\documentclass[twocolumn,english,showpacs,preprintnumbers,amsmath,amssymb,floatfix]{revtex4}
\usepackage[T1]{fontenc}
\usepackage[latin9]{inputenc}
\usepackage{color}
\usepackage{array}
\usepackage{amstext}
\usepackage{graphicx}
\usepackage{esint}

\makeatletter

\@ifundefined{textcolor}{}
{%
 \definecolor{BLACK}{gray}{0}
 \definecolor{WHITE}{gray}{1}
 \definecolor{RED}{rgb}{1,0,0}
 \definecolor{GREEN}{rgb}{0,1,0}
 \definecolor{BLUE}{rgb}{0,0,1}
 \definecolor{CYAN}{cmyk}{1,0,0,0}
 \definecolor{MAGENTA}{cmyk}{0,1,0,0}
 \definecolor{YELLOW}{cmyk}{0,0,1,0}
 }

\@ifundefined{definecolor}
 {\usepackage{color}}{}
\@ifundefined{definecolor}
 {\usepackage{color}}{}
\makeatother

\makeatother

\usepackage{babel}

\begin{document}

\title{Angular analysis of polarized top quark decay into B-mesons in two different helicity systems}

\author{S. Mohammad Moosavi Nejad$^{a,b}$}
\email{mmoosavi@yazd.ac.ir}

\author{Mahboobe Balali$^a$}

\affiliation{$^{(a)}$Faculty of Physics, Yazd University, P.O. Box
89195-741, Yazd, Iran}

\affiliation{$^{(b)}$School of Particles and Accelerators,
Institute for Research in Fundamental Sciences (IPM), P.O.Box
19395-5531, Tehran, Iran}

\date{\today}

\begin{abstract}

We calculate the ${\cal O}(\alpha_s)$ radiative corrections to the spin dependent differential decay rates
of the process $t\rightarrow b+W^+$. These are needed to study the angular distribution of the energy of
hadrons produced in polarized top quark decays at next-to-leading order (NLO).
In our previous work, we studied the angular distribution of the scaled-energy 
of bottom-flavored hadrons (B) from polarized top quark 
decays, using a specific helicity coordinate system where
the top quark spin was measured relative to the bottom momentum (system~1). Here, we study the angular distribution 
of the energy spectrum of B-hadron in a different helicity system, where the top spin is specified relative to 
the W-momentum (system~2). These energy distributions are governed by the  polarized and unpolarized
rate functions which are related to the density matrix elements of the decay $t\rightarrow W^++b$.
Through this paper, we present our predictions of the B-hadron spectrum in the polarized and unpolarized top decay and shall compare
the polarized results in two different helicity systems. 
These predictions can be used to determine the polarization states of top quarks and also provide direct access to the 
B-hadron fragmentation functions (FFs) and allow us to deepen our knowledge of the hadroniazation process.
\end{abstract}

\maketitle

\section{Introduction}
\label{sec:intro}

The top quark as a heaviest elementary particle, is the electroweak isospin partner 
of the bottom quark. Since its discovery  by the
CDF and $D0$ experiments at Fermilab Tevatron\cite{Group:2009ad},   
the determination of its properties has been one of the main goals of
the Tevatron Collider, recently joined by the CERN Large Hadron Collider (LHC).
The experiments at the  LHC will allow one to perform improved measurements of the top
properties, such as its  mass $m_t$ and branching fractions to high accuracy.
The measurement of the top mass, as a fundamental parameter of the standard
model (SM), has received particular attention. Indeed, the mass of top, the $W$-boson mass, and the Higgs boson mass 
are related through radiative corrections that provide an internal consistency check of the SM.
In a recent paper \cite{Abazov:2014dpa}, the mass of the top quark is measured as $m_t=174.98\pm 0.76$ GeV, by 
using the full sample of $p\bar{p}$ collision data collected by the $D0$ experiment in Run II of the Fermilab Tevatron Collider 
at $\sqrt{s}=1.96$ TeV. The theoretical aspects of top quark physics at the LHC are listed in \cite{Bernreuther:2008ju}.\\
The SM result of the top quark life time is $\tau_t\approx 0.5\times 10^{-24}$ s \cite{Chetyrkin:1999ju}
which is much shorter than the typical time for the formation of QCD bound states
$\tau_{QCD}\approx 1/\Lambda_{QCD}\approx3\times 10^{-24}$ s, i.e. the top quark
decays so rapidly that it does not have enough time to hadronize. Due to the Cabibbo-Kobayashi-Maskawa (CKM) mixing matrix 
element $V_{tb}\approx 1$ \cite{Cabibbo:1963yz}, the decay width of the top quark is dominated by the two-body 
channel $t\rightarrow b+W^+$ in the minimal SM of particle physics.
At the top mass scale the strong coupling constant is small, $\alpha_s(m_t)\approx 0.1$, so that the QCD effects involving
the top quark are well behaved in the perturbative sense. This allows one to apply the top quark decay as an 
appropriate tool for studying perturbative QCD and thus top decays provide a very clean source of information about the structure of the SM.\\
On the other hand, bottom quarks produced in the top decays hadronize before they decay and  the bottom 
hadronization ($b\rightarrow B+X$) is indeed one of the sources of uncertainty in the measurement of the top mass at 
the LHC \cite{M.Beneke} and the Tevatron \cite{Abulencia:2005ak}, as it contributes to the Monte Carlo systematics.
At the LHC, recent studies \cite{Kharchilava:1999yj} have suggested that final states with leptons, coming 
from the $W^+$ decay ($W^+\rightarrow l^+\nu_l$), and $J/\psi$, coming from the decay of a bottom-flavored  
meson (B), would be a promising channel to reconstruct the top mass.
At the LHC, of particular interest is the distribution in the scaled-energy of B-meson ($x_B$) in the top quark rest frame
as reliably as possible, so that this $x_B$ distribution provides direct access to the B-hadron fragmentation functions (FFs).
In \cite{Kniehl:2012mn}, in addition to the $x_B$ distribution, we also studied
the doubly differential partial width $d^2\Gamma/(dx_Bd\cos\theta)$ of the decay chain $t\rightarrow bW^+\rightarrow Bl^+\nu_l+X$,
where $\theta$ is the decay angle of the lepton in the W-boson rest frame. The $\cos\theta$ distribution allows one to analyze 
the $W^+$-boson polarization and so to further constrain the B-meson FFs.
In \cite{MoosaviNejad:2011yp}, we  studied the QCD NLO corrections to 
the energy distribution of B-mesons from  the decay of an unpolarized top quark into a stable charged-Higgs boson, 
$t\rightarrow BH^++X$, in the theories beyond-the-SM with an extended Higgs sector.
Although, in \cite{Ali:2011qf} it is mentioned that there is a clear separation between the decays $t \rightarrow bW^+$ and $t \rightarrow bH^+$ 
at the LHC, in both the $t\bar{t}X$ pair production and the  $t/\bar{t}X$ single top production.

The interplay between the top mass and its spin is of crucial importance in studying the SM.
Due to the top large mass, the top quark decays rapidly so that its life time scale is much shorter than the typical time required for the QCD 
interactions to randomize its spin, therefore its full spin information is preserved in the decay and passes on to its decay products. Hence, 
the top quark polarization  can be studied through the angular correlations between the direction of the top quark spin and the 
momenta of the decay products.  
Therefore, the particular purpose of this paper is to evaluate the QCD NLO corrections to the energy distribution of 
B-hadrons from the decay of a polarized top quark into a bottom quark, via $t(\uparrow)\rightarrow W^++b(\rightarrow B+X)$.
We mention that highly polarized top quarks will become available at hadron colliders
through single top production processes, which occur at the $33\%$ level of the $t\bar{t}$ pair production rate \cite{Mahlon:1996pn},
and in top quark pairs produced in future linear $e^+-e^-$-colliders \cite{Kuhn:1983ix}.
In \cite{Nejad:2013fba}, we studied the angular distribution of the  scaled-energy of the B/D-hadrons at NLO by calculating 
the polar angular correlation in the rest frame decay of a polarized top quark into a stable $W^+$-boson and 
B/D-hadrons, via $t(\uparrow)\rightarrow W^++D/B+X$. We analyzed this angular correlation in a special helicity coordinate system with
the event plane defined in the $(x, z)$ plane and the z-axes along the b-quark momentum. In this frame (system~1), the top quark polarization vector
was evaluated with respect to the direction of the b-quark momentum. Generally, to define the planes one needs to measure the momentum
directions of the momenta $\vec{p}_b$ and $\vec{p}_W$ and the polarization direction of the top quark, where the measurement of the momentum direction
of $\vec{p}_b$ requires the use of a jet finding algorithm, whereas the polarization direction of the top quark must be obtained from the theoretical input.
In electron-positron interactions the polarization degree of the top quark can be tuned with the help of polarized beams \cite{Parke},
so that a polarized linear electron-positron collider may be viewed as a copious source of close to zero and close to $100\%$ polarized top quarks.

In the present work, we analyze the angular distribution of the B-hadron energy 
in a different helicity coordinate system where, as before, the event plane is the $(x, z)$ plane but with the z-axes along the $W^+$-boson.
The polarization direction of the top quark is evaluated w.r.t this axes.
This coordinate system (system~2) makes the calculations more complicated because of the presence of the $W^+$-momentum $|\vec{p}_W|$ in the 
${\cal O}(\alpha_s)$ real amplitude of the process $t\rightarrow b+W^+$. To obtain the scaled distribution of B-hadron energy, at first we present an analytical 
expression for the NLO corrections to the differential width of the decay process $t(\uparrow)\rightarrow b+W^+$ in two different 
helicity coordinate systems and then using the
realistic and nonperturbative $b\rightarrow B$ FF we shall present and compare our results in both systems.
The measurement of  the  energy distribution of the B-hadron will be important to deepen our understanding of the 
nonperturbative aspects of B-hadrons  formation  and to test the universality and scaling violations of the B-hadron FFs while the 
angular analysis of the polarized top decay constrain these FFs even further.

This paper is structured as follows.
In Sec.~\ref{sec:angular}, we introduce the angular structure of differential decay widths. 
In Secs.~\ref{sec:born}-\ref{sec:nlo}, we present our analytic results for the angular distributions of partial decay rates 
in two different helicity systems at the Born level and next-to-leading order 
by introducing the technical details of our calculations.
In Sec.~\ref{sec:hadron}, we present our numerical analysis in hadron level and
in Sec.~\ref{sec:conclusion},  our conclusions are summarized.

\section{Angular structure of differential decay rate}
\label{sec:angular}

The dynamics of the current-induced $t\rightarrow b$ transition is embodied in the hadron tensor 
$H^{\mu\nu}\propto \sum_{X_b}\left\langle t(p_t, s_t)|J^{\mu\dagger}|X_b\right\rangle\left\langle X_b|J^\nu|t(p_t, s_t)\right\rangle$, where 
the SM  current combination  is given by $J_\mu=(J_\mu^V-J_\mu^A)\propto\bar\psi_b \gamma_\mu(1-\gamma_5)\psi_t$,
and $s_t$ stands for the top quark spin. 
Here, the intermediate states are $|X_b>=|b(p_b, s_b)>$ for the Born term and virtual one-loop contributions
and $|X_b>=|b+g>$ for the ${\cal O}(\alpha_s)$ real contributions.

In the rest frame of a top quark decaying into a b-quark, a $W^+$-boson and a gluon, the final state particles $b, W^+$ and gluon
define an event frame. Relative to this event plane one can define the polarization direction
of the polarized top quark. There are two various choices of possible coordinate systems relative to the event
plane where one differentiates between helicity systems according to the orientation of the $z$-axis. These systems are shown 
in Figs.~\ref{nlo1} (system~1) and \ref{nlo2} (system~2). In the system~1, the three-momentum of the b-quark
points into the direction of the positive $z$-axis and in the system~2, the momentum of the $W$-boson is defined
along the positive $z$-axis. 

Generally, the angular distribution of the differential decay width $d\Gamma/dx$ of a polarized top quark is expressed  by the following
 form  to clarify the correlation between the polarization of the top quark and its decay products  
\begin{eqnarray}\label{form}
\frac{d^2\Gamma}{dx_id\cos\theta_P}=\frac{1}{2}\bigg\{\frac{d\Gamma_A}{dx_i}+P\frac{d\Gamma_B}{dx_i}\cos\theta_P\bigg\},
\end{eqnarray}
where the polar angle $\theta_P$ shows the spin orientation of the top quark relative to the event plane and $P$ is the
magnitude of the top quark polarization. $P=0$ stands for an unpolarized top quark while $P=1$
corresponds to $100\%$ top quark polarization. In the notation above, $d\Gamma_A/dx$ and $d\Gamma_B/dx$ 
correspond to the unpolarized and polarized differential decay rates, respectively. 
As usual, we have defined the partonic scaled-energy fraction $x_i$ as
\begin{eqnarray}
x_i=\frac{2p_i\cdot p_t}{m_t^2}.
\end{eqnarray}
Neglecting the $b$-quark mass, one has $0\leq x_i\leq 1-\omega$ where $\omega$ is $\omega=m_W^2/m_t^2$.
Throughout this paper, we use the normalized partonic energy fraction as
\begin{eqnarray}\label{variable}
x_i=\frac{2E_i}{m_t(1-\omega)}, \qquad (i=b, g)
\end{eqnarray}
where $E_i$ stands for the energy of outgoing partons (bottom or gluon) and $0\leq x_i\leq 1$.

The ${\cal O}(\alpha_s)$ radiative corrections to the unpolarized differential rate $d\Gamma_A/dx$ have been 
studied in our previous work \cite{Kniehl:2012mn}, extensively.
The NLO radiative corrections to the polarized partial rate $d\Gamma_B/dx$ in the system~1 (Fig.~\ref{nlo1}) is studied in \cite{Nejad:2013fba} by one of us.
In the present work, we concentrate on the polarized top decay in the system~2 (Fig.~\ref{nlo2}) which is more complicated in comparison
with the analysis performed in the system~1. Finally, we shall compare our results in two coordinate systems~1 and 2 at the hadron level. 

\section{Born approximation}
\label{sec:born}

\begin{figure}
\begin{center}
\includegraphics[width=0.7\linewidth]{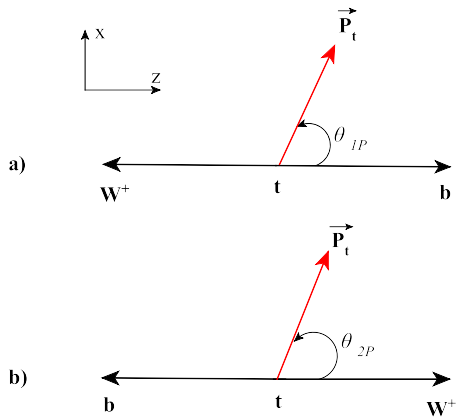}
\caption{\label{lo}%
Definition of the polar angle $\theta_{P}$ in two helicity systems. $\vec{P_t}$ is the top polarization vector.}
\end{center}
\end{figure}
It is straightforward to calculate the Born term contribution to the decay rate of the polarized top quark.
The Born term tensor is obtained from the square of the Born amplitude, given by
\begin{eqnarray}
M^{(0)}=V_{tb}\frac{g_{W}}{\sqrt{2}}\bar{u}_b\gamma^\mu\frac{1}{2}(1-\gamma_5)u_t,
\end{eqnarray}
where $g_W$ is related to the Fermi's constant $G_F$ as $g_W/\sqrt{2}=2m_W(G_F/\sqrt{2})^{1/2}$. After omitting the weak 
coupling factor $V_{tb}g_W/\sqrt{2}$ and summing over the b-quark spin, the Born term tensor reads
\begin{eqnarray}
B^{\mu\nu}&=&\frac{1}{4}Tr\{(\displaystyle{\not}p_b+m_b)\gamma^\mu(1-\gamma_5)(\displaystyle{\not}p_t+m_t)\times\nonumber\\
&&(1+\gamma_5\displaystyle{\not}s_t)\gamma^\nu(1-\gamma_5)\}.
\end{eqnarray}
Considering Fig.~\ref{lo}, we set the four-momentum and the polarization four-vector of the top quark as
\begin{eqnarray}
p_t=(m_t; \vec{0}), s_t=P(0; \sin\theta_P\cos\phi, \sin\theta_P\sin\phi, \cos\theta_P),\nonumber\\
\end{eqnarray}
and in the coordinate system~1 (Fig.~\ref{lo}a), the four-momentum of the b-quark  is set to $p_b=E_b(1; 0, 0, 1)$ and 
in the system~2 (Fig.~\ref{lo}b), it is $p_b=E_b(1; 0, 0, -1)$. Note that we put the b-quark mass to zero throughout this paper.
Therefore, the Born term helicity structure of differential rates in the system~1, reads
\begin{eqnarray}
\frac{d^2\Gamma_1^{\textbf{(0)}}}{dx_b d\cos\theta_{1P}}=\frac{1}{2}\bigg\{\Gamma_A^{\textbf{(0)}}-P\Gamma_B^{\textbf{(0)}}\cos\theta_{1P}\bigg\}\delta(1-x_b),
\end{eqnarray}
and in the system~2, is expressed as 
\begin{eqnarray}
\frac{d^2\Gamma_2^{\textbf{(0)}}}{dx_b d\cos\theta_{2P}}=\frac{1}{2}\bigg\{\Gamma_A^{\textbf{(0)}}+P\Gamma_B^{\textbf{(0)}}\cos\theta_{2P}\bigg\}\delta(1-x_b),
\end{eqnarray}
where,  $\Gamma_A^{\textbf{(0)}}$ corresponds to  the unpolarized Born term rate and $\Gamma_B^{\textbf{(0)}}$ describes the polarized Born rate.
They are given by
\begin{eqnarray}
\Gamma_A^{\textbf{(0)}}&=&\frac{\sqrt{2} m_t^3 G_F}{16\pi}(1+2\omega)(1-\omega)^2,\nonumber\\
\Gamma_B^{\textbf{(0)}}&=&\frac{\sqrt{2} m_t^3 G_F}{16\pi}(1-2\omega)(1-\omega)^2.
\end{eqnarray}
These results are in agreement with Refs.~\cite{Fischer:2001gp}, \cite{Fischer:1998gsa} and \cite{Groote:2006kq}. Setting $m_W=80.399$~GeV, 
$m_t=174.98$~GeV and $G_F=1.16637\times10^{-5}$~GeV$^{-2}$ one has $\Gamma_A^{\textbf{(0)}}=1.4335$ and $\Gamma_B^{\textbf{(0)}}=0.5939$.
Therefore, the polarization asymmetry $\alpha_W$, which is defined as $\alpha_W=\Gamma_B^{\textbf{(0)}}/\Gamma_A^{\textbf{(0)}}$, is $\alpha_W=0.396$.

\section{Virtual one-loop corrections}
\label{sec:virtual}

The required ingredients for the NLO calculation are the virtual one-loop contributions and the tree-graph
contributions. Since at the one-loop level, QED and QCD have the same structure then 
the virtual one-loop  corrections to the fermionic left-chiral (V-A) transitions have a long history,    
even dates back to QED times.\\ 
The virtual one-loop contributions into the polarized differential width are the same in
both helicity systems~1 and 2, and can be found in \cite{Nejad:2013fba}. 
We just mention that the virtual corrections arise from a virtual gluon exchanged between
the top and bottom quark legs (vertex correction), and from emission and absorption of a virtual
gluon from the same quark leg (quark self energy). Both of them include 
the IR and UV singularities, which are regularized by dimensional regularization 
in D space-time dimensions, where $D=4-2\epsilon$. All UV divergences are canceled after summing all
virtual contributions up, whereas the IR singularities are remaining, which are labeled by $\epsilon$ from now on.
Therefore, following the general form of the doubly differential distribution (\ref{form}), the virtual contribution in both coordinate systems is
\begin{eqnarray}\label{virtualfinal}
\frac{d^2\Gamma^{\textbf{vir}}}{dx_b d\cos\theta_P}=\frac{1}{2}\bigg\{\frac{d\Gamma^{\textbf{vir}}_A}{dx_b}+P\frac{d\Gamma^{\textbf{vir}}_B}{dx_b}\cos\theta_P\bigg\},
\end{eqnarray}
where 
\begin{eqnarray}
\frac{d\Gamma^{\textbf{vir}}_A}{dx_b}&=&\Gamma_A^{\textbf{(0)}}\frac{C_F\alpha_s}{2\pi}\bigg\{R-4\frac{1-\omega}{1-4\omega^2}\ln(1-\omega)\bigg\}\delta(1-x_b),\nonumber\\
\frac{d\Gamma^{\textbf{vir}}_B}{dx_b}&=&\Gamma_B^{\textbf{(0)}}\frac{C_F\alpha_s}{2\pi}R\delta(1-x_b).
\end{eqnarray}
In the equations above, $R$ is defined as 
\begin{eqnarray}
R&=&-\frac{F^2}{2}+\frac{F}{\epsilon}-\frac{1-4\omega}{1-2\omega}\ln(1-\omega)+2\ln\omega\ln(1-\omega)\nonumber\\
&&+2Li_2(1-\omega)-\frac{1}{\epsilon^2}-5\frac{\pi^2}{12}-\frac{23}{8},\nonumber\\
\end{eqnarray}
where, $F=2\ln(1-\omega)-\ln(4\pi\mu_F^2/m_t^2)+\gamma_E-\frac{5}{2}$.
Here, $\gamma_E=0.5772\cdots$ is the Euler Mascharoni constant, $Li_2(x)$ is the known dilogarithmic function and
$\mu_F$ is the QCD scale parameter.
The one-loop virtual contribution is purely real, as can be found from an inspection of the one-loop Feynman
diagrams, which does not accept any nonvanishing physical two-particle cut.

\section{QCD NLO contribution to angular distribution}
\label{sec:nlo}

At ${\cal O}(\alpha_s)$, the full amplitude of the transition $t\rightarrow b$ is the sum of the amplitudes of the Born term $M^{\textbf{(0)}}$, 
virtual one-loop $M^{\textbf{loop}}$, and the real gluon (tree-graph) contributions.
The real amplitude results from the decay $t(p_t)\rightarrow b(p_b)+W^+(p_W)+g(p_g)$, as
\begin{eqnarray}
M^{\textbf{real}}&=&e g_s\frac{T^n_{ij}}{2\sqrt{2}\sin\theta_W}\epsilon^{\star}_\beta(p_g, s_g)\epsilon_\mu(p_W, s_W)
\times\nonumber\\
&&\bar{u}(p_b, s_b)\bigg\{\gamma^\beta\frac{\displaystyle{\not}p_g+\displaystyle{\not}p_b}{2p_b.p_g}\gamma^\mu(1-\gamma_5)-\nonumber\\
&&\gamma^\mu(1-\gamma_5)\frac{m_t+\displaystyle{\not}p_t-\displaystyle{\not}p_g}{2p_t.p_g}\gamma^\beta\bigg\}u(p_t, s_t).
\end{eqnarray}
Here, $g_s=\sqrt{4\pi\alpha_s}$ is the strong coupling constant, $\theta_W$ is the weak mixing angle so that $\sin^2\theta_W=0.23124$ \cite{Nakamura:2010zzi},
and $"n"$ is the color index so $Tr(T^nT^n)/3=C_F$.
The polarization vectors of the gluon and the $W$-boson are also denoted by $\epsilon(p, s)$.\\
The QCD NLO contribution results from the square of the amplitudes as
$|M^{\textbf{(0)}}|^2$, $|M^{\textbf{vir}}|^2=2Re(M^{\textbf{(0)}\dagger} M^{\textbf{loop}})$ and $|M^{\textbf{real}}|^2=M^{\textbf{real} \dagger} M^{\textbf{real}}$.
To regulate the IR singularities, which arise from the soft- and collinear-gluon emission, we 
work in D-dimensions approach in which  to extract divergences we take the following replacement
\begin{eqnarray}
\int\frac{d^4p}{(2\pi)^4} \rightarrow \mu^{4-D}\int\frac{d^Dp}{(2\pi)^D},
\end{eqnarray}
where, $\mu$ is an arbitrary reference mass which shall be removed after summing all corrections up.
The differential decay rate for the real  contribution is given by
\begin{eqnarray}
d\Gamma^{\textbf{real}}=\frac{\mu_F^{2(4-D)}}{2m_t}\overline{|M^{\textbf{real}}|^2}dR_3(p_t, p_b, p_g, p_{W}),
\end{eqnarray}
where, the 3-body phase space element $dR_3$ reads
\begin{eqnarray}
\frac{d^{D-1}\bold{p}_b}{2E_b}\frac{d^{D-1}\bold{p}_W}{2E_W}\frac{d^{D-1}\bold{p}_g}{2E_g}
(2\pi)^{3-2D}\delta^D(p_t-\sum_{g,b,W} p_f).\nonumber\\
\end{eqnarray}

\begin{figure}
\begin{center}
\includegraphics[width=0.9\linewidth]{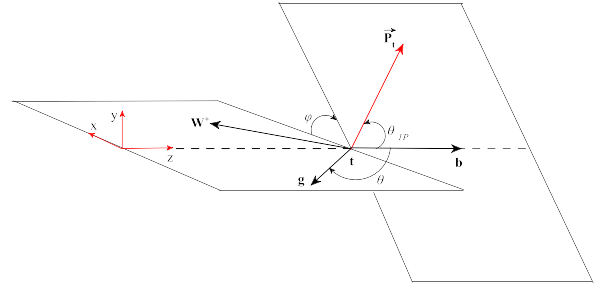}
\caption{\label{nlo1}%
Definition of the azimuthal angle $\phi$, the polarization vector of the top quark $\vec{P_t}$, 
the polar angles $\theta$ and $\theta_{P}$ in the helicity coordinate system~1.}
\end{center}
\end{figure}
\begin{figure}
\begin{center}
\includegraphics[width=0.9\linewidth]{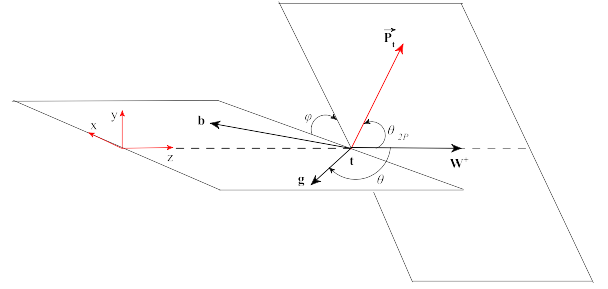}
\caption{\label{nlo2}%
As in Fig.~\ref{nlo1}, but in the helicity coordinate system~2.}
\end{center}
\end{figure}
To calculate the real doubly differential rate $d^2\Gamma^{\textbf{real}}/(dx_bd\cos\theta_P)$ and to 
get the correct finite terms, we normalize the polarized and the unpolarized doubly differential distributions to the 
corresponding Born widths evaluated in D-dimensions.
The polarized and unpolarized Born widths $\Gamma_B^{\textbf{(0)}}$ and $\Gamma_A^{\textbf{(0)}}$, evaluated in the 
dimensional regularization at ${\cal O}(\epsilon^2)$ are given
in Eq.~(29) of Ref.~\cite{Nejad:2013fba}.
Following Eq.~(\ref{form}), the ${\cal O}(\alpha_s)$ corrections to the angular distribution of partial decay rates 
are obtained by  summing the Born, the virtual and real gluon contributions and is given by
\begin{eqnarray}\label{first}
\frac{d^2\Gamma^{\textbf{nlo}}}{dx_b d\cos\theta_P}=\frac{1}{2}\bigg\{\frac{d\Gamma^{\textbf{nlo}}_A}{dx_b}+P\frac{d\Gamma^{\textbf{nlo}}_B}{dx_b}\cos\theta_P\bigg\}.
\end{eqnarray}
Generally, the contribution of the real gluon emission depends on the various 
choices of possible coordinate systems.
The results for $d\Gamma^{\textbf{nlo}}_A/dx_b$ are the same in both helicity systems and can be found in  \cite{Kniehl:2012mn}, 
and the analytical expression of the polarized angular distribution of decay width in the helicity system~1 ($d\Gamma^{\textbf{nlo}}_{1B}/dx_b$) is presented in \cite{Nejad:2013fba}.\\
To calculate the real differential rate $d\Gamma^{\textbf{real}}/dx_b$ in the coordinate system~2, 
we fix the momentum of b-quark and integrate over
the energy of the $W$-boson which ranges from $E_W^{\textbf{min}}=m_t(\omega+[1-x_b(1-\omega)]^2)/(2[1-x_b(1-\omega)])$ to  $E_W^{\textbf{max}}=m_t(1+\omega)/2$,
and to evaluate the angular distribution of differential width $d^2\Gamma^{\textbf{real}}/(dx_b d\cos\theta_{2P})$,
the angular integral in D-dimensions will have to be written as
\begin{eqnarray}
d\Omega_W=-\frac{2\pi^{\frac{D}{2}-1}}{\Gamma(\frac{D}{2}-1)}(\sin\theta_{2P})^{D-4} d\cos\theta_{2P}.
\end{eqnarray}
Therefore, the doubly differential distribution reads
\begin{eqnarray}\label{mohsen}
\frac{d^2\Gamma^{\textbf{real}}_2}{dx_b d\cos\theta_{2P}}&\propto& x_b^{D-4}\overline{|M^{real}|^2}(1-\cos^2\alpha)^{\frac{D-4}{2}}\times\nonumber\\
&&\delta(\cos\alpha-b)dE_W d\cos\alpha,
\end{eqnarray}
where, the coefficient of proportionality reads $\mu_F^{2(4-D)}(p_W m_t)^{D-4}(1-\omega)^{D-3}/(2^{3D-4}\pi^{D-1}\Gamma^2(\frac{D}{2}-1))$,
$b=(m_t^2+m_W^2-2m_t(E_b+E_W)+2E_bE_W)/(2E_bp_W)$, 
$p_W=\sqrt{E_W^2-m_W^2}$ is the momentum of $W$-boson and
$\alpha$ is the angle between the b-quark and the $W$-boson in Fig.~\ref{nlo2}.
Due to the presence of the W-momentum, working in the helicity system~2 is more complicated 
than the system~1 and for this reason our analytical results will not appear in a dinky form as in system~1 (see Eq.~(35) in \cite{Nejad:2013fba}).\\
Considering the top rest frame, the relevant scalar products evaluated in the system~2, are
\begin{eqnarray}
p_W\cdot s_t&=&-Pp_W\cos\theta_{2P},\quad p_b\cdot s_t=-PE_b\cos\alpha\cos\theta_{2P},\nonumber\\
&&\quad p_b\cdot p_W=E_b(E_W-p_W\cos\alpha),
\end{eqnarray}
and $p_t\cdot s_t=0$. Here, $P$ refers to the polarization degree of the top quark. 
To obtain the analytic result for the angular distribution of the differential rate at NLO, by summing the Born level, the virtual
and real gluon contributions, one has
\begin{eqnarray}\label{pol1}
\frac{d\Gamma^{\textbf{nlo}}_{2B}}{dx_b}&=&\Gamma_B^{\textbf{(0)}}\bigg\{\delta(1-x_b)+
\frac{C_F\alpha_s}{2\pi}\Big\{[-\frac{1}{\epsilon}+\gamma_E-\ln 4\pi]\times\nonumber\\
&&[\frac{3}{2}\delta(1-x_b)+\frac{1+x_b^2}{(1-x_b)_+}]+T_1\Big\}\bigg\},
\end{eqnarray}
where,
\begin{eqnarray}
	T_1&=&\delta(1-x_b)\big[\frac{3}{2}\ln\frac{m_t^2}{\mu_F^2}+2\ln\omega\ln(1-\omega)-\frac{2\pi^2}{3}+\nonumber\\
	&&\frac{2(1-\omega)}{1-2\omega}\ln(1-\omega)+4Li_2(1-\omega)-\nonumber\\
	&&\frac{2\omega}{1-\omega}\ln\omega-6\big]+2(1+x_b^2)\bigg(\frac{\ln(1-x_b)}{1-x_b}\bigg)_++\nonumber\\
	&&2\frac{1+x_b^2}{(1-x_b)_+}\ln[\frac{x_b(1-\omega)m_t}{\mu_F}]+\frac{1-S}{S(Sx_b^2-2 x_b+2)}\nonumber\\
	&&-\frac{2}{1-4S}-\frac{\omega}{S(1-x_b)}-1-x_b-\frac{1+x_b^2}{1-x_b}H_1\nonumber\\
	&&+\bigg(\frac{1}{S(1-x_b)}+\frac{4\omega}{(1-4S)(1-2Sx_b)}-\nonumber\\
	&&\frac{1-S}{S(Sx_b^2-2x_b+2)}\bigg)|2Sx_b^2-2x_b+1|\nonumber\\
	&&-\frac{H_2}{\sqrt{S(Sx_b^2-2x_b+2)}}\bigg(\frac{1}{S}+\frac{2}{1-4S}+\nonumber\\
	&&
	\frac{(13S-5-4S^2)x_b}{1-4S}-\frac{Sx_b(x_b^2-x_b+2)}{1-x_b}-\nonumber\\
	&&\frac{(1-S)[1-(1-S)x_b]}{S(Sx_b^2-2x_b+2)}\bigg),
\end{eqnarray}
where, $S=(1-\omega)/2$, $H_1=\ln\big[(1-S)x_b^2-x_b+\frac{1}{2}+\frac{1}{2}|2Sx_b^2-2x_b+1|\big]$ and
$H_2=\ln\big[(1-x_b)(1-3Sx_b)+Sx_b^2(1-2Sx_b)+\sqrt{S(Sx_b^2-2x_b+2)}
|2Sx_b^2-2x_b+1|\big]-\ln\big[1+(S-1)x_b+\sqrt{S(Sx_b^2-2x_b+2)}\big]$.\\
One can compare our polarized and unpolarized results against known results presented in \cite{Fischer:2001gp}.\\
Since the detected mesons in top decays can be also produced through  a fragmenting real gluon, therefore, to obtain 
the most accurate energy spectrum of B-meson we have to add the contribution of gluon fragmentation to 
the b-quark one to produce the outgoing meson. As shown in \cite{Nejad:2013fba},  this contribution can be 
important at a low energy of the observed meson so that this decreases the size of decay rate at the threshold. 
Therefore,  the angular distribution of the differential 
decay rate $d\Gamma/dx_g$ is also required,  where $x_g$ is defined in (\ref{variable}). 
Considering the general form of the angular distribution (\ref{form}), for the gluon contribution one has
\begin{eqnarray}
\frac{d^2\Gamma^{\textbf{nlo}}}{dx_g d\cos\theta_P}=\frac{1}{2}\bigg\{\frac{d\Gamma^{\textbf{nlo}}_A}{dx_g}+P\frac{d\Gamma^{\textbf{nlo}}_B}{dx_g}\cos\theta_P\bigg\},
\end{eqnarray}
where, the results for $d\Gamma^{\textbf{nlo}}_A/dx_g$ are the same in both coordinate systems and can be found in \cite{Kniehl:2012mn}, 
and the analytical expression for the polarized angular distribution in the helicity system~1 ($d\Gamma^{\textbf{nlo}}_{1B}/dx_g$) is presented in  \cite{Nejad:2013fba}.
In the system~2, to obtain the doubly differential distribution $d^2\Gamma/(dx_g d\cos\theta_{2P})$ we fix the momentum of the gluon and integrate over
the energy of $W$-boson which ranges from $E_W^{\textbf{min}}=m_t(\omega+[1-x_g(1-\omega)]^2)/(2[1-x_g(1-\omega)])$ to  $E_W^{\textbf{max}}=m_t(1+\omega)/2$.
Therefore, the doubly differential decay rate is given by
\begin{eqnarray}
\frac{d^2\Gamma^{\textbf{real}}_2}{dx_g d\cos\theta_{2P}}&\propto& x_g^{D-4}\overline{|M^{real}|^2}(1-\cos^2\theta)^{\frac{D-4}{2}}\times\nonumber\\
&&\delta(\cos\theta-a)dE_W d\cos\theta,
\end{eqnarray}
where, the proportionality coefficient is as in (\ref{mohsen}),  $\theta$ is the polar angle between the gluon 
and the $W$-boson (see Fig.~\ref{nlo2}), whereas $a=(m_t^2+m_W^2-2m_t(E_g+E_W)+2E_gE_W)/(2E_gp_W)$.
The relevant scalar products are 
\begin{eqnarray}
p_W\cdot s_t&=&-Pp_W\cos\theta_{2P},\quad p_g\cdot s_t=-PE_g\cos\theta\cos\theta_{2P},\nonumber\\
&& p_g\cdot p_W=E_g(E_W-p_W\cos\theta).
\end{eqnarray}
Therefore, in the coordinate system~2 the polarized differential width is expressed as
\begin{eqnarray}\label{pol2}
\frac{d\Gamma^{\textbf{nlo}}_{2B}}{dx_g}&=&
\Gamma_B^{\textbf{(0)}}\frac{C_F\alpha_s}{2\pi}\bigg\{\frac{1+(1-x_g)^2}{x_g}(-\frac{1}{\epsilon}+\gamma_E-\ln 4\pi)\nonumber\\
&&+T_2\bigg\},
\end{eqnarray}
where,
\begin{eqnarray}
T_2&=&\frac{1+(1-x_g)^2}{x_g}\bigg(-H_4+2\ln[x_g(1-\omega)(1-x_g)\frac{m_t}{\mu_F}]\bigg)\nonumber\\
&&\hspace{-0.8cm}+\bigg(-x_g+\frac{1+\omega^2}{4S^3x_g^2}+\frac{8S^2-6S+3}{S(4S-1)}+\frac{2-\omega^2(2\omega-5)}{2S^2(1-4S)x_g}\bigg)H_3\nonumber\\
&&\hspace{-0.8cm}+\bigg(\frac{S-1}{2S^2x_g^2}+\frac{\omega}{2(1-2Sx_g)^2}-\frac{12S^2-15S+7}{2S(1-4S)x_g}-\nonumber\\
&&\hspace{-0.8cm}\frac{24S^2-26S+9}{2(1-4S)(1-2Sx_g)}\bigg)|2Sx_g^2-2x_g+1|\nonumber\\
&&\hspace{-0.8cm}+x_g+\frac{1-S}{2S^2x_g^2}-\frac{2}{1-4S}+\frac{12S^2-7S+5}{2S(1-4S)x_g},
\end{eqnarray}
where $H_3=\ln\big[1-S(-2Sx_g^2+2x_g+1-|2Sx_g^2-2x_g+1|)\big]-\ln[1-2Sx_g]$ and $H_4=\ln\big[(1-S)x_g^2-x_g+\frac{1}{2}+\frac{1}{2}|2Sx_g^2-2x_g+1|\big]$.
In Eqs.~(\ref{pol1}) and (\ref{pol2}), $T_1$ and $T_2$ are free of all singularities and to subtract the collinear singularities 
remaining in the polarized partial widths, we apply the modified minimal-subtraction $(\overline{MS})$ scheme where, the singularities are absorbed
into the bare FFs. This renormalizes the FFs and creates the finite terms of the form $\alpha_s\ln(m_t^2/\mu_F^2)$ in 
the polarized differential widths. According to this scheme, to get the $\overline{MS}$ coefficient functions one shall has to subtract from
Eqs.~(\ref{pol1}) and (\ref{pol2}),  the ${\cal O}(\alpha_s)$ term multiplying the  characteristic $\overline{MS}$ constant $(-1/\epsilon+\gamma_E-\ln 4\pi)$. 
In the present work we set $\mu_F=m_t$, so that the terms proportional to $\ln(m_t^2/\mu_F^2)$ vanish. \\
We mention that the dimensional reduction scheme can be converted to the gluon mass regulator scheme by the 
replacement $1/\epsilon-\gamma_E+\ln(4\pi\mu_F^2/m_t^2)\rightarrow \ln\Lambda^2$, where $\Lambda=m_g/m_t$ is the scaled gluon mass.

\section{Angular distribution results in Hadron level}
\label{sec:hadron}

After determination of the differential decay rates in the parton level, we are now in a position to 
explore our phenomenological predictions of the energy distribution of B-meson 
by performing a numerical analysis in the two helicity coordinate systems.
In fact, we wish to calculate the quantity $d\Gamma/dx_B$, where the normalized energy
fraction of the outgoing meson is defined as $x_B=2E_B/(m_t(1-\omega))$, in similarity to  the parton level one (\ref{variable}).
The necessary tool to obtain the B-meson energy spectrum is the factorization  theorem of the QCD-improved parton model \cite{jc}, 
where the energy distribution of a hadron is expressed as the convolution of the parton-level 
spectrum with the nonperturbative FF $D_i^B(z, \mu_F)$
\begin{equation}
\frac{d\Gamma}{dx_B}=\sum_{i=b, g}\int_{x_i^\text{min}}^{x_i^\text{max}}
\frac{dx_i}{x_i}\,\frac{d\Gamma_i}{dx_i}(\mu_R, \mu_F) D_i(\frac{x_B}{x_i}, \mu_F),
\label{eq:master}
\end{equation}
where, $d\Gamma_i/dx_i$ is the partial width  of the parton-level process $t\to i(=b, g)+X$, with $X$ including  
the $W$-boson and any other parton. Here, $\mu_F$ and $\mu_R$ are the factorization and the renormalization scales, respectively, that
the scale $\mu_R$ is associated with the renormalization of the strong coupling constant and a normal 
choice, which we adopt in this work is $\mu_R=\mu_F=m_t$. In (\ref{eq:master}), $D_{i=b, g}(z, \mu_F)$  
is the nonperturbative FF of the transition $b/g\rightarrow B$ which is process independent.
It means, we can exploit data from $e^+e^-\rightarrow b\bar{b}$ processes to predict the b-quark hadronization in top decay. 
Note that the definitions of $d\Gamma/dx_i$ and $D_i^B(z, \mu_F)$ are not unique, but they depend on
the scheme which is used to subtract the collinear singularities appeared in the differential widths (\ref{pol1}) and (\ref{pol2}).
As we mentioned, in our work the $\overline{MS}$ factorization scheme is chosen.\\
Several models, including some fittable parameters have been already proposed to describe the nonperturbative transition from
a quark- to a hadron-state.
Following Ref.~\cite{Kniehl:2008zza}, we employ the B-hadron FFs determined at NLO in the zero-mass scheme,
through a global fit to $e^+e^-$-annihilation data presented by 
ALEPH \cite{Heister:2001jg} and OPAL \cite{Abbiendi:2002vt} collaborations at CERN LEP1 and by SLD \cite{Abe:1999ki} at SLAC SLC.
Specifically, at the initial scale $\mu_0=m_b$ the power model $D_b^B(z, \mu_0)=Nz^\alpha(1-z)^\beta$ is proposed for 
the $b\rightarrow B$ transition, while the gluon FF is set to zero and is  
evolved to higher scales using the Dokshitzer-Gribov-Lipatov-Alteralli-Parisi  (DGLAP) equations \cite{dglap}.
The results for the fit parameters are $N=4684.1,\alpha=16.87$ and $\beta=2.628$.  
As numerical input values, from \cite{Nakamura:2010zzi} we take
$G_F=1.16637\times10^{-5}$~GeV$^{-2}$, $m_W=80.339$ GeV,
$m_b=4.78$~GeV, 
$m_B=5.279$~GeV, and 
the typical QCD scale $\Lambda_{\overline{\text{MS}}}^{(5)}=231.0$~MeV adjusted such that $\alpha_s^{(5)}(m_Z=91.18)=0.1184$.
In the $\overline{\text{MS}}$ scheme the b-quark mass only enter through the initial condition of the  FF.\\
Before studying the B-hadron spectrum, we turn to our numerical results 
of the unpolarized and polarized decay rates in both helicity systems. In fact, we 
integrate $d\Gamma/dx_b$ (Eqs.~(\ref{pol1}), (35) from \cite{Nejad:2013fba} and (7) from \cite{Kniehl:2012mn})
over $x_b (0\leq x_b\leq 1)$, while the strong coupling constant is evolved from  $\alpha_s(m_Z)=0.1184$ to $\alpha_s(m_t)=0.1070$.
The normalized result for the polarized decay width in the helicity system~1 is
\begin{eqnarray}
\frac{\Gamma_{1B}^{\textbf{nlo}}}{\Gamma_B^{\textbf{(0)}}}=1-0.1303,
\end{eqnarray}
and for the one in the system~2, is
\begin{eqnarray}
\frac{\Gamma_{2B}^{\textbf{nlo}}}{\Gamma_B^{\textbf{(0)}}}=1-0.1162,
\end{eqnarray}
and the unpolarized decay rate normalized to the corresponding Born term,  is
\begin{eqnarray}
\frac{\Gamma_A^{\textbf{nlo}}}{\Gamma_A^{\textbf{(0)}}}=1-0.08542.
\end{eqnarray}
To study the $x_B$ scaled-energy  distributions of B-hadrons
produced in the polarized top decay, we consider the quantity
$d\Gamma(t(\uparrow)\to B+X)/dx_B$ in the two helicity coordinate systems.
In \cite{Kniehl:2012mn,Nejad:2013fba}, we showed that the $g\rightarrow B$ contribution into the NLO energy spectrum of the 
B-meson is negative and appreciable only
in the low-$x_B$ region and for higher values of $x_B$ the NLO result is
practically exhausted by the $b\rightarrow B$ contribution. The contribution of the gluon is calculated to see where
it contributes to $d\Gamma/dx_B$ and can not be discriminated in the meson spectrum  as an experimental quantity.
\begin{figure}
\begin{center}
\includegraphics[width=0.8\linewidth,bb=37 22 452 440]{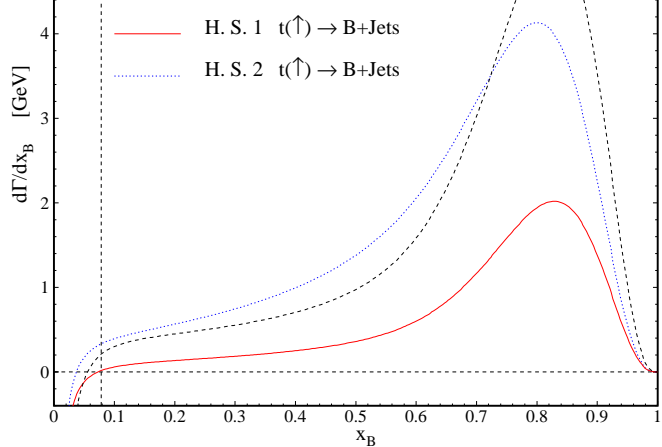}
\caption{\label{figure1}%
$d\Gamma_B^{\textbf{nlo}}/dx_B$ as a function of $x_B$ in the helicity system~1 (solid line)
and the system~2 (dotted line). The polarized results are also compared to the unpolarized one $d\Gamma_A^{\textbf{nlo}}/dx_B$ (dashed line).
Threshold at $x_B$ is shown.}
\end{center}
\end{figure}

\begin{figure}
\begin{center}
\includegraphics[width=0.8\linewidth,bb=37 22 452 440]{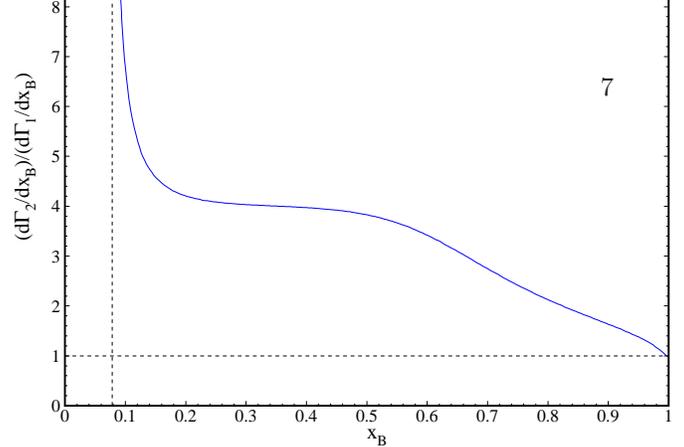}
\caption{\label{figure2}%
$x_B$ spectrum at NLO in the helicity system~2, normalized to the one in the  system~1.}
\end{center}
\end{figure}
In Fig.~\ref{figure1}, the $x_B$-spectrum of the B-hadron produced in the unpolarized  top quark decay (dashed line) is shown.
The polarized ones in the helicity system~1 (solid line) and system~2 (dotted line) are also studied. 
As is seen, the differential decay width of the polarized top in the helicity system~2 (H.~S.~2) is totally higher than the one
in the helicity system~1 (H.~S.~1).  
For a more quantitative interpretation of Fig.~\ref{figure1}, we consider in Fig.~\ref{figure2} the
partial decay width $d\Gamma_B^{\textbf{nlo}}/dx_b$ in the H.~S.~2 normalized to the one in the H.~S.~1.
Note that  all results are valid just for $x_B\geq 2m_B/(m_t(1-\omega))=0.078$.

\section{Conclusion}
\label{sec:conclusion}

Studying the fundamental properties of the top quark is one of the main fields of investigation
in theoretical and experimental particle physics.
The short life time of the top quark implies that it decays before hadronization takes
place; therefore, it retains its full polarization content 
and passes on the spin information to its decay products. 
This allows us to study the top-spin state using the angular distributions of its decay products.
Whereas the bottom quark, produced through the top decay, hadronizes therefore
the distributions in the B-hadron energy  are of
particular interest. In \cite{Kniehl:2012mn}, we studied the scaled-energy distribution
of the B-meson in unpolarized top quark decays $t\rightarrow W^++b(\rightarrow B)$. In \cite{Nejad:2013fba}, we
made our predictions for the scaled-energy distributions of the B- and D-mesons from polarized top decays
using a special helicity coordinate system, 
where the event plane lies in the $(x, z)$ plane and the bottom momentum is along the $z$-axis.
In the present work, we have presented results on the 
${\cal O}(\alpha_s)$ radiative corrections to the spin dependent differential width $d^2\Gamma/(dx_B d\cos\theta_P)$,
applying a different helicity system where the $z$-axis is defined by the $W$-momentum.
This provides independent probes of the polarized top quark decay dynamics.
To obtain these results we presented, for the first time, the analytical results for the parton-level 
differential decay widths of $t\rightarrow b+W^+$ in two helicity systems and then
we compared our results in both systems. We found that the polarized results depend on the
selected helicity system, extremely.\\
On one hand, the $x_B$ distributions provide direct access to the B-hadron FFs, and on the other hand
the universality and scaling violations of the B-hadron FFs will be able to test at LHC by comparing
our predictions with future measurements of $d\Gamma/dx_B$.
The $\cos\theta_P$ distribution allows one to analyze the polarization state of top quarks, where the polar angle $\theta_P$ refers 
to the angle between the top polarization vector  and the $z$-axis.
The formalism made here is also applicable to the other hadrons such as pions and
kaons, using the $(b, g)\rightarrow (\pi, K)$ FFs which can be found in \cite{maryam}.

\begin{acknowledgments}
We would like to thank Professor B.~A.~Kniehl for reading  the manuscript and also for his important comments.
S.~M.~Moosavi Nejad thanks the CERN
TH-PH division for its hospitality, where a portion of this
work was performed. Thanks to  Z.~Hamedi for reading and improving the English
manuscript.
\end{acknowledgments}

\end{document}